\documentclass[11pt]{cernrep}
\usepackage{amssymb}
\usepackage{graphicx}
\usepackage{cite,./mcite}

\begin{document}

\title{PQCD Formulations with Heavy Quark Masses and Global Analysis\thanks{%
~~Contribution to the HERA - LHC Workshop Proceedings }}
\author{Robert S Thorne$^1$ and W.K. Tung$^2$.}
\institute{\rule{0em}{3ex}$^1$ University College London, {}$^2$ Michigan State
University and University of Washington}
\maketitle

\begin{abstract}
We critically review heavy quark mass effects in DIS and their impact on
global analyses. We lay out all elements of a properly defined general mass
variable flavor number scheme (GM VFNS) that are shared by all modern
formulations of the problem. We then explain the freedom in choosing
specific implementations and spell out, in particular, the current
formulations of the CTEQ and MSTW groups. We clarify the approximations in
the still widely-used zero mass variable flavor scheme (ZM VFNS), mention
the inherent flaws in its conventional implementation, and consider the
possibility of mending some of these flaws. We discuss practical issues
concerning the use of parton distributions in various physical applications,
in view of the different schemes. And we comment on the possible presence of
intrinsic heavy flavors.
\end{abstract}

\tableofcontents


\institute{ {}$^1$ University College London, {}$^2$ Michigan State U. and
U. Washington, Seattle}

\newpage

\section{Introduction}

The proper treatment of heavy flavours in global QCD analysis of parton
distribution functions (PDFs) is essential for precision measurements at
hadron colliders. Recent studies \cite{cteq65,MSTW,cteq66,Watt08} show that
the standard-candle cross sections for $W/Z$ production at the LHC are
sensitive to detailed features of PDFs that depend on heavy quark mass
effects; and certain standard model as well as beyond standard model
processes depend crucially on better knowledge of the $c$-quark parton
density, in addition to the light parton flavors. These studies also make it
clear that the consistent treatment of heavy flavours in perturbative QCD
(PQCD) require theoretical considerations that go beyond the familiar
textbook parton picture based on massless quarks and gluons. There are
various choices, explicit and implicit, which need to be made in various
stages of a proper calculation in generalised PQCD including heavy quark
mass effects. In the global analysis of PDFs, these choices can affect the
resulting parton distributions. Consistent choices are imperative; mistakes
may result in differences that are similar to, or even greater than, the
quoted uncertainties due to other sources (such as the propagation of input
experimental errors). In this report, we will provide a brief, but full,
review of issues related to the treatment of heavy quark masses in PQCD,
embodied in the general mass variable flavor scheme (GM VFNS).

In Sec.\thinspace \ref{sec:GenFrm}, we describe the basic features of the
modern PQCD formalism incorporating heavy quark masses. In Sec.\thinspace %
\ref{sec:implement}, we first delineate the common features of GM VFNS, then
identify the different (but self-consistent) choices that have been made in
recent global analysis work, and compare their results. For readers
interested in practical issues relating to the use (or choice) of PDFs in
various physics applications, we present a series of comments in Sec.~\ref%
{sec:use} intended as guidelines. In Sec.~\ref{sec:IC}, we discuss the
possibility of intrinsic heavy flavors.

We note that, this review on GM VFNS and global analysis is not intended to
address the specific issues pertinent to heavy flavor production (especially
the final state distributions). For this particular process, somewhat
different considerations may favor the adoption of appropriate fixed flavor
number schemes (FFNS). We shall not go into details of these considerations;
but will mention the FFNS along the way, since the GM VFNS is built on a
series of FFNS's. We will comment on this intimate relationship whenever
appropriate.

\section{General Considerations on PQCD with Heavy Flavor Quarks\label%
{sec:GenFrm}}

The quark-parton picture is based on the factorization theorem of PQCD. The
conventional proof of the factorization theorem proceeds from the zero-mass
limit for all the partons---a good approximation at energy scales
(generically designated by $Q$) far above all quark mass thresholds
(designated by $m_{i}$). This clearly does not hold when $Q/m_{i}$ is of
order 1.\footnote{%
Heavy quarks, by definition, have $m_{i}\gg \Lambda _{QCD}$. \ Hence we
always assume $Q,m_{i}\gg \Lambda _{QCD}$. In practice, $i=c,b,t$.} It has
been recognised since the mid-1980's that a consistent treatment of heavy
quarks in PQCD over the full energy range from $Q\lesssim m_{i}$ to $Q\gg
m_{i}$ can be formulated \cite{ColTun}. In 1998, Collins gave a general
proof of the factorization theorem (order-by-order to all orders of
perturbation theory) that is valid for non-zero quark masses \cite{collins}.
The resulting general theoretical framework is conceptually simple: it
represents a straightforward generalisation of the conventional zero-mass
(ZM) modified minimal subtraction ($\overline{\mathrm{MS}}$) formalism and
it contains the conventional approaches as special cases in their respective
regions of applicability; thus, it provides a good basis for our discussions.

The implementation of any PQCD calculation on physical cross sections
requires attention to a number of details, both kinematical and dynamical,
that can affect both the reliability of the predictions. Physical
considerations are important to ensure that the right choices are made
between perturbatively equivalent alternatives that may produce noticeable
differences in practical applications. It is important to make these
considerations explicit, in order to make sense of the comparison between
different calculations in the literature. This is what we shall do in this
section. \ In subsequent sections, we shall point out the different choices
that have been made in recent global analysis efforts.

Heavy quark physics at HERA involve mostly charm ($c$) and bottom ($b$)
production; at LHC, top ($t$) production, in addition, is of interest. For
simplicity, we often focus the discussion of the theoretical issues on the
production of a single heavy quark flavor, which we shall denote generically
as $H$, with mass $m_{H}$. The considerations apply to all three cases, $%
H=c,~b,~\&~t$. For global analysis, the most important process that requires
precision calculation is DIS; hence, for physical predictions, we will
explicitly discuss the total inclusive and semi-inclusive structure
functions, generically referred to as $F^{\lambda }(x,Q)$, where $\lambda $
represents either the conventional label ($1,2,3$) or the alternative ($%
T,L,3 $) where $T/L$ stands for transverse/longitudinal respectively.

\subsection{The Factorization Formula}

The PQCD factorization theorem for the DIS structure functions has the
general form
\begin{equation}
F_{\lambda }(x,Q^{2})=\sum_{k}f_{k}\otimes C_{k}^{\lambda
}=\sum_{k}\int_{\chi }^{1}{\frac{d\xi }{\xi }}\ f_{k}(\xi ,\mu )\
C_{k}^{\lambda }\left( \frac{\chi }{\xi },\frac{Q}{\mu },\frac{m_{i}}{\mu }%
,\alpha _{s}(\mu )\right) .  \label{master}
\end{equation}%
Here, the summation is over the active parton flavor label $k$, $f^{k}(x,\mu
)$ are the parton distributions at the factorization scale $\mu $, $%
C_{k}^{\lambda }$ are the Wilson coefficients (or hard-scattering
amplitudes) that can be calculated order-by-order in perturbation theory.
The lower limit of the convolution integral $\chi $ is determined by
final-state phase-space constraints: in the conventional ZM parton formalism
it is simply $x=Q^{2}/2q\cdot p$---the Bjorken $x$---but this is no longer
true when heavy flavor particles are produced in the final state, cf.\
Sec.\thinspace \ref{sec:rescaling} below. \ The renormalization and
factorization scales are jointly represented by $\mu $:\ in most
applications, it is convenient to choose $\mu =Q$; but there are
circumstances in which a different choice becomes useful.

\subsection{Partons and Schemes for General Mass PQCD\label{sec:suma}}

In PQCD, the summation $\sum_{k}$ over ``parton flavor'' label $k$ in the
factorization formula, Eq.\thinspace (\ref{master}), is determined by the
\emph{factorization scheme} chosen to \emph{define} the Parton Distributions
$f_{k}(x,\mu )$.

If mass effects of a heavy quark $H$ are to be taken into account, the
simplest scheme to adopt is the \emph{fixed flavor number scheme} (FFNS) in
which all quark flavors below $H$ are treated as zero-mass and one sums over
$k=g,u,\bar{u},d,\bar{d},...$ up to $n_{f}$ flavors of \emph{light}
(massless) quarks. The mass of $H$, $m_{H}$, appears explicitly in the
Wilson coefficients \{$C_{k}^{\lambda }$\}, as indicated in Eq.\,\ref{master}%
. For $H=\{c,b,t\}$, $n_{f}=\{3,4,5\}$ respectively. Historically,
higher-order ($\mathcal{O}(\alpha_S^2)$) calculations of the heavy quark
production \cite{NLOFFNS} were all done first in the FFNS. These
calculations provide much improved results when $\mu $ ($Q$) is of the order
of $m_{H}$ (both above and below), over those of the conventional ZM ones
(corresponding to setting $m_{H}=0$).

Unfortunately, at any finite order in perturbative calculation, the $n_{f}$%
-FFNS results become increasingly unreliable as $Q$ becomes large compared
to $m_{H}$: the Wilson coefficients contain logarithm terms of the form $%
\alpha _{s}^{n}\ln ^{m}(Q/m_{H})$, where $m=1\ldots n$, at order $n$ of the
perturbative expansion, implying they are not infrared safe---higher order
terms do not diminish in size compared to lower order ones---the
perturbative expansion eventually breaks down. \ Thus, even if all $n_{f}$%
-flavor FFNS are mathematically equivalent, in practice, the 3-flavor scheme
yields the most reliable results in the region $Q\lesssim m_{c}$, the
4-flavor scheme in $m_{c}\lesssim Q\lesssim m_{b}$, the 5-flavor scheme in $%
m_{b}\lesssim Q\lesssim m_{t}$, and, if needed, the 6-flavor scheme in $%
m_{t}\lesssim Q\ $. (Cf.\thinspace related discussions later in this
section.)

This leads naturally to the definition of the more general \emph{variable
flavor number scheme} (VFNS): it is a \emph{composite scheme} consists of
the sequence of $n_{f}$-flavor FFNS, each in its region of validity, for $%
n_{f}=3,4,..$ as described above; and the various $n_{f}$-flavor schemes are
related to each other by perturbatively calculable transformation
(finite-renormalization) matrices among the (running) coupling $\alpha _{s}$%
, the running masses $\left\{ m_{H}\right\} $, the parton distribution
functions $\left\{ f_{k}\right\} $, and the Wilson coefficients $%
\{C_{k}^{\lambda }\}$. These relations ensure that there are only one set of
independent renormalization constants, hence make the definition of the
composite scheme precise for all energy scale $\mu ~(Q)$; and they ensure
that physical predictions are well-defined and continuous as the energy
scale traverses each of the overlapping regions $Q\sim m_{H}$ where both the
$n_{f}$-flavor and the $(n_{f}+1)$-flavor schemes are applicable. The
theoretical foundation for this intuitively obvious scheme can be found in
\cite{ColTun,collins}, and it was first applied in detail for structure
functions in \cite{ACOT}. Most recent work on heavy quark physics adopt this
general picture, in one form or another. \ We shall mention some common
features of this general-mass (GM) VFNS in the next few paragraphs; and
defer the specifics on the implementation of this scheme, as well as the
variations in the implementation allowed by the general framework until
Sec.\,\ref{sec:implement}.

As mentioned above, the $n_{f}$-flavor and the $(n_{f}+1)$-flavor schemes
within the GM VFNS should be matched at some \emph{match point} $\mu _{M}$
that is of the order of $m_{H}$. \ In practice, the matching is commonly
chosen to be exactly $\mu _{M}=m_{H}$, since it has been known that, in the
calculational scheme appropriate for GM VFNS\footnote{%
Technically, this means employing the CWZ subtraction scheme \cite{CWZ} in
calculating the higher-order Feynman diagrams. CWZ subtraction is an elegant
extension of the $\overline{\mathrm{MS}}$ subtraction scheme that ensures
the decoupling of heavy quarks at high energy scales order-by-order. This is
essential for factorization to be valid at each order of perturbation
theory. (In the original $\overline{\mathrm{MS}}$ subtraction scheme,
decoupling is satisfied only for the full perturbation series---to infinite
orders.)\label{fn:cwz}}, the transformation matrices vanish at this
particular scale at NLO in the perturbative expansion \cite{ColTun}; thus
discontinuities of the renormalized quantities are always of higher order,
making practical calculations simpler in general.

Strictly speaking, once the component $n_{f}$-flavor schemes are
unambiguously matched, one can still choose an independent \emph{transition
scale}, $\mu _{T}$, at which to switch from the $n_{f}$-flavor scheme to the
$(n_{f}+1)$-flavor scheme in the calculation of physical quantities in
defining the GM VFNS. This scale must again be within the overlapping
region, but can be different from $\mu _{M}$ \cite{collins,cteq65}. In fact,
it is commonly known that, from the physics point of view, in the region
above the $m_{H}$ threshold, up to $\eta \,m_{H}$ with a reasonable-sized
constant factor $\eta $, the most natural parton picture is that of $n_{f}$%
-flavor, rather than $(n_{f}+1)$-flavor one.\footnote{%
Specifically, the $n_{f}$-flavor scheme should fail when $\alpha
_{s}(\mu )\ln (\mu /m_{H})=\alpha _{s}(\mu )\ln (\eta )$ ceases to be a
small parameter for the effective perturbation expansion. However, no
theory can tell us precisely how small is acceptably \textquotedblleft
small\textquotedblright ---hence how large $\eta $ is permitted. Ardent
FFNS advocates believe even the range of the 3-flavor scheme extends to
all currently available energies, including HERA \cite{GRV}. For GM
VFNS, see the next paragraph. \label{fn:transition}} For instance, the
$3$-flavor scheme calculation has been favored by most HERA work on
charm and bottom quark production, even if the HERA DIS kinematic
region mostly involves $Q>m_{c}$; and it is also used in the
dynamically generated parton approach to global analysis \cite{GRV}.

In practice, almost all implementations of the GM VFNS simply choose $\mu
_{T}=\mu _{M}=m_{H}$ (often not explicitly mentioning the conceptual
distinction between $\mu _{T}$ and $\mu _{M}=m_{H}$). The self-consistency
of the GM VFNS guarantees that physical predictions are rather insensitive
to the choice of the transition point as long as it is within the
overlapping region of validity of the $n_{f}$- and $(n_{f}+1)$-flavor ones.
The simple choice of $\mu _{T}=m_{H}$ corresponds to opting for the lower
end of this region for the convenience in implementation. In the following,
we shall use the terms matching point\ and transition point interchangeably.
As with all definition ambiguities in perturbative theory, the sensitivity
to the choice of matching and transition points diminishes at higher orders.

\subsection{Treatment of Final-state Flavors\label{sec:sumb}}

For total inclusive structure functions, the factorization formula,
Eq.\thinspace (\ref{master}), contains an implicit summation over all
possible quark flavors in the final state. One can write,%
\begin{equation}
C_{k}=\sum_{j}C_{k}^{j}  \label{Wilson}
\end{equation}%
where \textquotedblleft $j$\textquotedblright\ denotes final state flavors,
and $\{C_{k}^{j}\}$ represent the Wilson coefficients (hard cross sections)
for an incoming parton \textquotedblleft $k$\textquotedblright\ to produce a
final state containing flavor \textquotedblleft $j$\textquotedblright\
calculable perturbatively from the relevant Feynman diagrams. It is
important to emphasize that \textquotedblleft $j$\textquotedblright\ labels
quark flavors that can be produced \emph{physically} in the final state; it
is \emph{not} a \emph{parton} label in the sense of initial-state parton
flavors described in the previous subsection. The latter (labeled $k$) is a
theoretical construct and scheme-dependent (e.g.\thinspace it is fixed at
three for the 3-flavor scheme); whereas the final-state sum (over $j$) is
over \emph{all flavors} that can be physically produced. Furthermore, the
initial state parton \textquotedblleft $k$\textquotedblright\ does not have
to be on the mass-shell, and is commonly treated as massless; whereas the
final state particles \textquotedblleft $j$\textquotedblright\ should
certainly be \emph{on-mass-shell} in order to satisfy the correct kinematic
constraints for the final state phase space and yield physically meaningful
results.\footnote{%
Strict kinematics would require putting the produced heavy flavor mesons or
baryons on the mass shell. In the PQCD formalism, we adopt the approximation
of using on-shell final state heavy quarks in the underlying partonic
process.} Thus, in implementing the summation over final states, the most
relevant physical scale is $W$---the CM energy of the virtual Compton
process---in contrast to the scale $Q$ that controls the initial state
summation over parton flavors.

The distinction between the two summations is absent in the simplest
implementation of the conventional (i.e., textbook) zero-mass parton
formalism: if all quark masses are set to zero to begin with, then all
flavors can be produced in the final state. \ This distinction becomes
blurred in the commonly used zero-mass (ZM) VFNS, where the heavy quark
masses \{$m_{H}$\} implicitly enter because the number of effective parton
flavors is incremented as the scale parameter $\mu $ crosses each heavy
quark threshold. \ This creates apparent paradoxes in the implementation of
the ZM VFNS, such as: for $\mu =Q<m_{b}$, $b$ is not counted as a parton,
the partonic process $\gamma +g\rightarrow b\bar{b}$ would not be included
in DIS calculations, yet physically this can be significant if $W\gg 2m_{b}$
(small $x$); whereas for $\mu =Q>m_{b}$, $b$ is counted as a massless
parton, the contribution of $\gamma +g\rightarrow b\bar{b}$ to DIS would be
the same as that of $\gamma +g\rightarrow d\bar{d}$, but physically this is
wrong for moderate values of $W$, and furthermore, it should be zero if $%
W<2m_{b}$ (corresponding to large $x$). (We shall return to this topic in
Sec.\thinspace \ref{sec:improveZM}.)

These problems were certainly overlooked in conventional global analyses
from its inception until the time when issues on mass-effects in PQCD were
brought to the fore after the mid 1990's \cite{ACOT,transitionme1,
transitionme2,trvfns1,trvfns2}. Since then, despite its shortcomings the
standard ZM VFNS continues to be used widely because of its simplicity and
because NLO Wilson coefficients for most physical processes are still only
available in the ZM VFNS. Most groups produce the standard ZM VFNS as either
their default set or as one of the options, and they form the most common
basis for comparison between groups, e.g. the ``benchmark study'' in \cite%
{HERALHC}.

It is obvious that, in a proper implementation of PQCD with mass (in any
scheme), the distinction between the initial-state and final-state summation
must be unambiguously, and correctly, observed. \ For instance, even in the
3-flavor regime (when $c$ and $b$ quarks are \emph{not counted as partons}),
the charm and bottom flavors still need to be counted in the final
state---at tree-level via $W^{\mathrm{+}}+d/s\rightarrow c$, and at 1-loop
level via the gluon-fusion processes such as $W^{\mathrm{+}}+g\rightarrow
\bar{s}+c$ or $\gamma +g\rightarrow c\bar{c}\,(b\bar{b})$, provided there is
enough CM energy to produce these particles.

\subsection{Phase-space Constraints and Rescaling\label{sec:rescaling}}

The above discussion points to the importance of the proper treatment of
final state phase space in heavy quark calculations. Once mass effects are
taken into account, kinematic constraints have a significant impact on the
numerical results of the calculation; in fact, they represent the dominant
factor in the threshold regions of the phase space. \ In DIS, with heavy
flavor produced in the final state, the simplest kinematic constraint that
comes to mind is
\begin{equation}
W-M_{N}>\sum_{f}~M_{f}  \label{KinConstraint}
\end{equation}%
where $W$ is the CM energy of the vector-boson--nucleon scattering process, $%
M_{N}$ is the nucleon mass, and the right-hand side is the sum of \emph{all}
masses in the final state. $W$ is related to the familiar kinematic
variables ($x,Q$) by $W^{2}-M_{N}^{2}=Q^{2}(1-x)/x$, and this constraint
should ideally be imposed 
on the right-hand side of Eq.\thinspace (\ref{master}).
Any approach achieving this represents an improvement over the conventional
ZM scheme calculations, that ignores the kinematic constraint Eq.\thinspace (%
\ref{KinConstraint}) (resulting in a gross over-estimate of the
corresponding cross sections). The implementation of the constraint in the
most usual case of NC processes, say $\gamma /Z+c\rightarrow c$ (or any
other heavy quark) is not automatic (and is absent in some earlier
definitions of a GM VFNS) because in this partonic process one must account
for the existence of a \emph{hidden heavy particle}---the $\bar{c}$---in the
target fragment. The key observation is, heavy objects buried in the target
fragment are still a part of the final state, and should be included in the
phase space constraint, Eq.\thinspace (\ref{KinConstraint}).

Early attempts to address this issue were either approximate or rather
cumbersome, and could not be naturally extended to high orders.\footnote{%
In \cite{ACOT}, the threshold violation was minimized by an artificial
choice of the factorization scale $\mu (m_{H},Q)$. In \cite{trvfns1,trvfns2}
the kinematic limit was enforced exactly by requiring continuity of the
slope of structure functions across the matching point, resulting in a
rather complicated expression for the coefficient functions in Eq.(\ref%
{master}).} A much better physically motivated approach is based on the idea
of rescaling. The simplest example is given by charm production in the LO CC
process $W+s\rightarrow c$. It is well-known that, when the final state
charm quark is put on the mass shell, kinematics requires the momentum
fraction variable for the incoming strange parton, $\chi $ in Eq.\thinspace (%
\ref{master}) to be $\chi =x(1+m_{c}^{2}/Q^{2})$ \cite{Barnett76}, rather
than the Bjorken $x$. This is commonly called the \emph{rescaling variable}.
The generalization of this idea to the more prevalent case of NC processes
took a long time to emerge \cite{ACOTchi1,ACOTchi2} which extended the
simple rescaling to the more general case of $\gamma /Z+c\rightarrow c+X$,
where $X$ contains only light particles, it was proposed that the
convolution integral in Eq.\thinspace (\ref{master}) should be over the
momentum fraction range $\chi _{c}<\xi <1$, where%
\begin{equation}
\chi _{c}=x\left( 1+\frac{4m_{c}^{2}}{Q^{2}}\right) \ \ .  \label{rescaling}
\end{equation}%
In the most general case where there are any number of heavy particles in
the final state, the corresponding variable is (cf.\thinspace Eq.\thinspace (%
\ref{KinConstraint}))
\begin{equation}
\chi =x\left( 1+\frac{\left( \Sigma _{f}~M_{f}\right) ^{2}}{Q^{2}}\right) \
\ .  \label{Rescaling}
\end{equation}%
This rescaling prescription has been referred to as ACOT$\chi $ in the
recent literature \cite{ACOTchi1,ACOTchi2,nnlovfns}.

Rescaling shifts the momentum variable in the parton distribution function $%
f^{k}(\xi ,\mu )$ in Eq.\thinspace (\ref{master}) to a higher value than in
the zero-mass case. For instance, at LO, the structure functions $F_{\lambda
}(x,Q)$ are given by some linear combination of $f^{k}(x,Q)$ in the ZM
formalism; but, with ACOT$\chi $ rescaling, this becomes $f^{k}(\chi _{c},Q)$%
. In the region where $\left( \Sigma _{f}\,M_{f}\right) ^{2}/Q^{2}$ is not
too small, especially when $f(\xi ,\mu )$ is a steep function of $\xi $,
this rescaling can substantially change the numerical result of the
calculation. \ It is straightforward to show that, when one approaches a
given threshold ($M_{N}+\Sigma _{f}~M_{f}$) from above, the corresponding
rescaling variable $\chi \rightarrow 1$. Since generally $f^{k}(\xi ,\mu
)\longrightarrow 0$ as $\xi \rightarrow 1$, rescaling ensures a smoothly
vanishing threshold behavior for the contribution of the heavy quark
production term to all structure functions. This results in a universal%
\footnote{%
Since it is imposed on the (universal) parton distribution function part of
the factorization formula.}, and intuitively physical, realization of the
threshold kinematic constraint for all heavy flavor production processes
that is applicable to all orders of perturbation theory. \ For this reason,
most recent global analysis efforts choose this method.

\subsection{Difference between \{$F_{\protect\lambda }^\mathrm{tot}$\} and \{%
$F_{\protect\lambda }^\mathrm{H}$\} Structure Functions\label{sec:F2FH}}

In PQCD, the most reliable calculations are those involving infra-red safe
quantities---these are free from logarithmic factors that can become large
(thereby spoiling the perturbative expansion). The total inclusive structure
functions $\{F_{\lambda }^\mathrm{tot}\}$ defined in the GM VFNS are
infrared safe, as suggested by the discussion of Sec.\,\ref{sec:suma} and
proven in Ref.\,\cite{collins}.

Experimentally, the semi-inclusive DIS structure functions for producing a
heavy flavor particle in the final state is also of interest. Theoretically,
it is useful to note that the structure functions \{$F_{\lambda }^{\mathrm{H}%
}$\} for producing heavy flavor $H$ are not as well defined as $F_{\lambda
}^{\mathrm{tot}}$.\footnote{%
In the following discussion, we shall overlook logarithmic factors normally
associated with fragmentation functions for simplicity. These are similar to
those associated with parton distributions, but are less understood from the
theoretical point of view---e.g.\thinspace the general proof of
factorization theorem (with mass) \cite{collins} has not yet been extended
to cover fragmentation.} To see this, consider the relation between the two,%
\begin{equation}
F_{\lambda }^{\mathrm{tot}}=F_{\lambda }^{\mathrm{light}}+F_{\lambda }^{%
\mathrm{H}}\ ,  \label{totalinc}
\end{equation}%
where $F_{\lambda }^{\mathrm{light}}$ denotes the sum of terms with only
light quarks in the final state, and $F_{\lambda }^{\mathrm{H}}$ consist of
terms with at least one heavy quark $H$ in the final state. \ Unfortunately,
$F_{\lambda }^{\mathrm{H}}(x,Q,m_{H})$ is, strictly speaking, \emph{not
infrared safe} beyond order $\alpha _{s}$ (1-loop): they contain residual $%
\ln ^{n}(Q/m_{H})$ terms at higher orders (2-loop and up). The same terms
occur in $F_{\lambda }^{\mathrm{light}}$ due to contributions from virtual $%
H $ loops, with the opposite sign. Only the sum of the two, i.e.\thinspace
the total inclusive quantities $F_{\lambda }^{\mathrm{tot}}$ are infra-red
safe. 
This problem could be addressed properly by adopting a physically motivated,
infrared-safe cut-off on the invariant mass of the heavy quark pair,
corresponding to some experimental threshold \cite{Chuvakin} in the
definition of $F_{\lambda }^{\mathrm{H}}$ (drawing on similar practises in
jet physics). In practice, up to order $\alpha _{s}^{2}$, the result is
numerically rather insensitive to this, and different groups adopt a variety
of less sophisticated procedures, e.g. including contributions with virtual $%
H$ loops within the definition of $F_{\lambda }^{\mathrm{H}}$. Nonetheless,
it is prudent to be aware that the theoretical predictions on $F_{\lambda }^{%
\mathrm{H}}$ are intrinsically less robust than those for $F_{\lambda }^{%
\mathrm{tot}}$ when comparing experimental results with theory calculations.

\subsection{Conventions for ``\textrm{LO}'' , ``\textrm{NLO}'' , ...
Calculations\label{sec:order}}

It is also useful to point out that, in PQCD, the use of familiar terms such
as LO, NLO, ... is often ambiguous, depending on which type of physical
quantities are under consideration, and on the convention used by the
authors. This can be a source of considerable confusion when one compares
the calculations of $F_{\lambda }^\mathrm{tot}$ and$\ F_{\lambda }^\mathrm{H}
$ by different groups (cf.\,next section).

One common convention is to refer LO results as those derived from tree
diagrams; NLO those from 1-loop calculations, ... and so on. This convention
is widely used; and it is also the one used in the CTEQ papers. \ Another
possible convention is to refer to LO results as the \emph{first non-zero
term} in the perturbative expansion; NLO as one order higher in $\alpha _{s}$%
, ... and so on. This convention originated in FFNS calculations of heavy
quark production; and it is also used by the MRST/MSTW authors. It is a
process-dependent convention, and it depends \textit{a priori} on the
knowledge of results of the calculation to the first couple of orders in $%
\alpha _{s}$.

Whereas the two conventions coincide for quantities such as $F_{2}^{\mathrm{%
tot}}$; they lead to different designations for the longitudinal structure
function $F_{L}^{\mathrm{tot}}$ and the $n_{f}$-flavor $F_{2}^{\mathrm{%
H,n_{f}}}$, since the tree-level results are zero for these quantities.
These designations, by themselves, are only a matter of terminology.
However, mixing the two distinct terminologies in comparing results of
different groups can be truly confusing. \ This will become obvious later.

\section{Implementations of VFNS: Common Features and Differences\label%
{sec:implement}}

In this section, we provide some details of the PQCD basis for the GM VFNS,
and comment on the different choices that have been made in the various
versions of this general framework, implemented by two of the major groups
performing global QCD analysis.

\subsection{Alternative Formulations of the ZM\ VFNS}

\label{sec:improveZM}

As pointed out in Sec.\thinspace \ref{sec:sumb}, the ZM VFNS, as commonly
implemented, represents an unreliable approximation to the correct PQCD in
some kinematic regions because of inappropriate handling of the final-state
counting and phase-space treatment, in addition to the neglect of
heavy-quark mass terms in the Wilson coefficients. \ Whereas the latter is
unavoidable to some extent, because the massive Wilson coefficients have not
yet been calculated even at 1-loop level for most physical processes (except
for DIS), the former (which can be more significant numerically in certain
parts of phase space) can potentially be remedied by properly counting the
final states and using the rescaling variables, as discussed in
Secs.\thinspace \ref{sec:sumb} and \thinspace \ref{sec:rescaling} under
general considerations. Thus, alternative formulations of the ZM VFNS are
possible that only involve the zero-mass approximation in the Wilson
coefficient. This possibility has not yet been explicitly explored.\

\subsection{Parton Distribution Functions in VFNS (ZM and GM)}

In PQCD, the factorization scheme is determined by the choices made in
defining the parton distribution functions (as renormalized Green
functions). In a GM VFNS based on the generalized $\overline{\mathrm{MS}}$
subtraction (cf.\thinspace footnote \ref{fn:cwz}) the evolution kernel of
the DGLAP equation is \emph{mass-independent}; thus the PDFs, so defined,
apply to GM VFNS calculations as they do for the ZM VFNS.

In the VFNS, the PDFs switch from the $n_{f}$-flavor FFNS ones to the $%
(n_{f}+1)$-flavor FFNS ones at the matching point $\mu =m_{H}$ (cf.\,Sec.\,%
\ref{sec:suma}); the PDFs above/below the matching point are related,
order-by-order in $\alpha _{s}$, by:
\begin{equation}
f_{j}^{VF}(\mu \rightarrow m_{H}^{\mathrm{+}})\equiv
f_{j}^{(n_{f}+1)FF}=A_{jk}\otimes f_{k}^{n_{f}FF}\equiv A_{jk}\otimes
f_{j}^{VF}(\mu \rightarrow m_{H}^{\mathrm{-}}),  \label{transition}
\end{equation}%
where $m_{H}^{+/-}$ indicate that the $\mu \rightarrow m_{H}$ limit is taken
from above/below, and we have used the shorthand VF/FF for VFNS/FFNS in the
superscripts. The transition matrix elements $A_{jk}(\mu /m_{H})$,
representing a finite-renormalization between the two overlapping FFNS
schemes, can be calculated order by order in $\alpha _{s}$; they are known
to NNLO, i.e. $\mathcal{O}(\alpha _{S}^{2})$ \cite%
{transitionme1,transitionme2}. (Note that $A_{jk}$ is not a square matrix.)
It turns out, at NLO, $A_{jk}(\mu =m_{H})=0 $ \cite{collins}; thus $%
f_{k}^{VF}$ are continuous with this choice of matching point. There is a
rather significant discontinuity in heavy quark distributions and the gluon
distribution at NNLO.

With the matching conditions, Eq.\,\ref{transition}, \{$f_{j}^{VF}(\mu )$\}
are uniquely defined for all values of $\mu $. We shall omit the superscript
VF in the following. Moreover, when there is a need to focus on $f_{j}(\mu )$
in the vicinity of $\mu =m_{H}$, where there may be a discontinuity, we use $%
f_{j}^{+/-}(\mu )$ to distinguish the above/below branch of the function. As
indicated in Eq.\,\ref{transition}, $f_{j}^\mathrm{-}$ correspond to the $%
n_{f}$-flavor PDFs, and $f_{j}^\mathrm{+}$ to the $(n_{f}+1)$-flavor ones.

\subsection{The Structure of a GM VFNS, Minimal Prescription and Additional
Freedom}

Physical quantities should be independent of the choice of scheme; hence, in
a GM VFNS, we must require the theoretical expressions for the structure
functions to be continuous across the matching point $\mu =Q=m_{H}$ to each
order of perturbative theory:
\begin{eqnarray}
F(x,Q)=C_{k}^{\mathrm{-}}(m_{H}/Q)\otimes f_{k}^{\mathrm{-}}(Q) &=&C_{j}^{%
\mathrm{+}}(m_{H}/Q)\otimes f_{j}^{\mathrm{+}}(Q)  \label{C-match} \\
&\equiv &C_{j}^{\mathrm{+}}(m_{H}/Q)\otimes A_{jk}(m_{H}/Q)\otimes f_{k}^{%
\mathrm{-}}(Q).  \label{matchF}
\end{eqnarray}%
where we have suppressed the structure function label ($\lambda $) on $F$'s
and $C$'s, and used the notation $C_{k}^{+/-}$ to denote the Wilson
coefficient function $C_{k}(m_{H}/Q)$ above/below the matching point
respectively. Hence, the GM VFNS coefficient functions are also, in general,
discontinuous, and must satisfy the transformation formula:
\begin{equation}
C_{k}^{\mathrm{-}}(m_{H}/Q)=C_{j}^{\mathrm{+}}(m_{H}/Q)\otimes
A_{jk}(m_{H}/Q).  \label{VFNSdef}
\end{equation}%
order-by-order in $\alpha _{s}$. \ For example, at $\mathcal{O}(\alpha _{S})$%
, $A_{Hg}=\alpha _{s}P_{qg}^{0}\ln (Q/m_{H})$, this constraint implies,
\begin{equation}
C_{H,g}^{\mathrm{-},1}(m_{H}/Q)=\alpha _{s}C_{H,H}^{\mathrm{+}%
,0}(m_{H}/Q)\otimes P_{qg}^{0}\ln (Q/m_{H})+C_{H,g}^{\mathrm{+},1}(m_{H}/Q).
\label{LOVFNSdef}
\end{equation}%
where the numeral superscript (0,1) refers to the order of calculation in $%
\alpha _{s}$ (for $P_{jk}$, the order is by standard convention one higher
then indicated), and the suppressed second parton index on the Wilson
coefficients (cf.\thinspace Eq.\thinspace \ref{Wilson}) has been restored to
make the content of this equation explicit. Eq.\thinspace (\ref{LOVFNSdef})
was implicitly used in defining the original ACOT scheme \cite{ACOT}. The
first term on the RHS of Eq.\thinspace \ref{LOVFNSdef}, when moved to the
LHS, becomes the \emph{subtraction term} of Ref.\thinspace \cite{ACOT} that
serves to define the Wilson coefficient $C_{H,g}^{\mathrm{+},1}(m_{H}/Q)$
(hence the scheme) at order $\alpha _{s}$, as well as to eliminate the
potentially infra-red unsafe logarithm in the gluon fusion term ($C_{H,g}^{%
\mathrm{-},1}(m_{H}/Q)$) at high energies.

The GM VFNS as described above, consisting of the general framework of \cite%
{ColTun,collins}, along with transformation matrices \{$A_{jk}$\} calculated
to order $\alpha _{s}^{2}$ by \cite{transitionme1,transitionme2}, is
accepted in principle by all recent work on PQCD with mass. \ Together, they
can be regarded as the \emph{minimal GM VFNS}.

The definition in Eq.\thinspace \ref{VFNSdef} was applied to find the
asymptotic limits ($Q^{2}/M_{H}^{2}\rightarrow \infty $) of coefficient
functions in \cite{transitionme1,transitionme2}, but it is important to
observe that it does not completely define all Wilson coefficients across
the matching point, hence, there are additional flexibilities in defining a
specific scheme \cite{trvfns1,trvfns2,collins,SACOT}. This is because, as
mentioned earlier, the transition matrix \{$A_{jk}$\} is not a square
matrix---it is $n_{f}\times (n_{f}+1)$. It is possible to swap $\mathcal{O}%
(m_{H}/Q)$ terms between Wilson coefficients on the right-hand-side of
Eq.\thinspace \ref{VFNSdef} (hence redefining the scheme) without violating
the general principles of a GM VFNS. For instance, one can swap $\mathcal{O}%
(m_{H}/Q)$ terms between $C_{H}^{+,0}(m_{H}/Q)$ and $C_{g}^{\mathrm{+}%
,1}(m_{H}/Q)$ while keeping intact the relation (\ref{LOVFNSdef}) that
guarantees the continuity of $F(x,Q)$ according to Eq.\thinspace \ref%
{C-match}. This general feature, applies to (\ref{VFNSdef}) to all orders.
It means, in particular, that there is no need to calculate the coefficient
function $C_{H}^{\mathrm{+},i}(m_{H}/Q)$, for any $i$ -- it can be chosen as
a part of the definition of the scheme. Also, it is perfectly possible to
define coefficient functions which do not individually satisfy the
constraint in Eq.\thinspace \ref{KinConstraint}, since Eq.\thinspace \ref%
{VFNSdef} guarantees ultimate cancellation of any violations between terms.
However, this will not occur perfectly at any finite order so modern
definitions do include the constraint explicitly, as outlined in
Sec.\thinspace \ref{sec:rescaling}.

The additional flexibility discussed above has been exploited to simply the
calculation, as well as to achieve some desirable features of the prediction
of the theory by different groups. Of particular interest and usefulness is
the general observation that, given a GM VFNS calculation of \{$C_{j}^{%
\mathrm{+}}$\}, one can always switch to a simpler scheme with constant \{$%
\tilde{C}_{j}^{\mathrm{+}} $\}
\begin{equation}
\tilde{C}_{H}^{\mathrm{+}}(m_{H}/Q)=C_{H}^{\mathrm{+}}(0)  \label{sacot}
\end{equation}%
This is because the shift ($C_{H}^{\mathrm{+}}(m_{H}/Q)-C_{H}^{\mathrm{+}%
}(0) $) vanishes in the $m_{H}/Q\rightarrow 0$ limit, and can be absorbed
into a redefinition of the GM scheme as mentioned above. The detailed proof
are given in \cite{collins,SACOT}. By choosing the heavy-quark-initiated
contributions to coincide with the ZM formulae, the GM VFNS calculation
becomes much simplified: given the better known ZM results, we only need to
know the full $m_{H}$-dependent contributions from the
light-parton-initiated subprocesses; and these are exactly what is provided
by the $n_{f}$-flavor FFNS calculations available in the literature. This
scheme is known as the \emph{Simplified ACOT scheme}, or SACOT~\cite%
{collins,SACOT}.

Further uses of the freedom to reshuffle $\mathcal{O}(m_{H}/Q)$ terms
between Wilson coefficients, as well as adding terms of higher order in the
matching condition (without upsetting the accuracy at the given order) have
been employed extensively by the MRST/MSTW group, as will be discussed in
Sec.\,\ref{sec:mrst}.

\subsection{CTEQ Implementation of the GM VFNS\label{sec:cteq}}

The CTEQ group has always followed the general PQCD framework as formulated
in \cite{ColTun,collins}. Up to CTEQ6.1, the default CTEQ PDF sets were
obtained using the more familiar ZM Wilson coefficients, because, the vast
majority of HEP applications carried out by both theorists and
experimentalists use this calculational scheme. For those applications that
emphasized heavy quarks, special GM VFNS PDF sets were also provided; these
were named as CTEQnHQ, where $n=4,5,6$.

The earlier CTEQ PDFs are now superseded by CTEQ6.5 \cite{cteq65} and
CTEQ6.6 \cite{cteq66} PDFs; these are based on a new implementation of the
general framework described in previous sections, plus using the simplifying
SACOT choice of heavy quark Wilson coefficients \cite{ACOT,KreBein}
specified by Eq.\thinspace \ref{sacot} above. There are no additional
modification of the formulae of the minimal GM VFNS, as described in
previous sections. CTEQ uses the convention of designating tree-level,
1-loop, 2-loop calculations as LO, NLO, and NNLO, for all physical
quantities, $F_{\lambda }^{\mathrm{tot}}$, $F_{\lambda }^{\mathrm{H}}$, ...
etc., cf.\ Sec.\thinspace \ref{sec:order}.

With these minimal choices, this implementation is extremely simple.
Continuity of physical predictions across matching points in the scale
variable $\mu =Q $ is guaranteed by Eqs.\,\ref{C-match} and \ref{VFNSdef};
and continuity across physical thresholds in the physical variable $W$, for
producing heavy flavor final states, are guaranteed by the use of ACOT-$\chi
$ rescaling variables \ref{Rescaling}, as described in Sec.\,\ref%
{sec:rescaling}.

For example, to examine the continuity of physical predictions to NLO in
this approach, we have, for the below/above matching point calculations:
\begin{equation}
\begin{array}{rcl}
F_{2}^{-H}(x,Q^{2}) & = & \alpha _{s}C_{2,Hg}^{-,1}\otimes g^{n_{f}} \\
F_{2}^{+H}(x,Q^{2}) & = & \alpha _{s}C_{2,Hg}^{+,1}\otimes
g^{n_{f}+1}+(C_{2,HH}^{+,0}+\alpha _{s}C_{2,HH}^{+,1})\otimes (h+\bar{h})%
\end{array}
\label{ACOTNLO}
\end{equation}%
where non-essential numerical factors have been absorbed into the
convolution $\otimes $. The continuity of $F_{2}^{H}(x,Q^{2})$ in the
scaling variable $\mu =Q$ is satisfied by construction (Eq.\,\ref{matchF})
because the relation between the PDFs given by Eq.\,\ref{transition} and
that between the Wilson coefficients given by Eq.\,\ref{C-match} involve the
same transformation matrix \{$A_{jk}$\} (calculated in \cite%
{transitionme1,transitionme2,Chuvakin}). In fact, to this order, $%
A_{Hg}=\alpha _{s}P_{qg}^{0}\ln (Q/m_{H})$, hence
\begin{eqnarray*}
h(\bar{h}) &=&0 \\
g^{n_{f}+1} &=&g^{n_{f}} \\
C_{2,Hg}^{+,1} &=&C_{2,Hg}^{-,1}~,
\end{eqnarray*}
at the matching point $\mu =Q=m_{H}$. Thus, the two lines in Eq.\,\ref%
{ACOTNLO} give the same result, and $F_{2}^{H}(x,Q^{2})$ is continuous. The
separate issue of continuity of $F_{2}^{H}(x,Q^{2})$ in the physical
variable $W$ across the production threshold of $W=2m_{H}$ is satisfied
automatically by each individual term (using the ACOT-$\chi $ prescription
for the quark terms and straightforward kinematics for the gluon term).

In the CTEQ approach, all processes are treated in a uniform way; there is
no need to distinguish between neutral current (NC) and charged current (CC)
processes in DIS, (among others, as in MRST/MSTW). All CTEQ global analyses
so far are carried out up to NLO. This is quite adequate for current
phenomenology, given existing experimental and other theoretical
uncertainties.\ Because NNLO results has been known to show signs of
unstable behavior of the perturbative expansion, particularly at small-$x$,
they are being studied along with resummation effects that can stabilize the
predictions. This study is still underway.

\subsection{MRST/MSTW Implementation of the GM VFNS\label{sec:mrst}}

\subsubsection{Prescription\label{sec:mrstF2}}

In the TR heavy flavour prescriptions, described in \cite{trvfns1,trvfns2}
the ambiguity in the definition of $C_{2,HH}^{\mathrm{VF},0}(Q^{2}/m_{H}^{2})
$ was exploited by applying the constraint that $(d\,F_{2}^{\mathrm{H}%
}/d\,\ln Q^{2})$ was continuous at the transition point (in the gluon
sector). However, this becomes technically difficult at higher orders.
Hence, in \cite{nnlovfns} the choice of heavy-flavour coefficient functions
for $F_{2}^{H}$ was altered to be the same as the SACOT($\chi $) scheme
described above. This choice of heavy-flavour coefficient functions has been
used in the most recent MRST/MSTW analysis, in the first instance in \cite%
{MSTW}. To be precise the choice is
\begin{equation}
C_{2,HH}^{\mathrm{VF},n}(Q^{2}/m_{H}^{2},z)=C_{2,HH}^{\mathrm{ZM}%
,n}(z/x_{max}).
\end{equation}%
This is applied up to NNLO in \cite{nnlovfns} and in subsequent analyses.
For the first time at this order satisfying the requirements in Eq.(\ref%
{VFNSdef}) leads to discontinuities in coefficient functions, which up to
NNLO cancel those in the parton distributions. This particular choice of
coefficient functions removes one of the sources of ambiguity in defining a
GM VFNS. However, there are additional ambiguities in the MRST/MSTW
convention for counting LO, NLO, ...\ calculations (cf.\thinspace Sec.\ref%
{sec:order}), coming about because the ordering in $\alpha _{S}$ for $%
F_{2}^{H}(x,Q^{2})$ is different above and below matching points in
Eqs.\thinspace \ref{matchF}-\ref{LOVFNSdef}. (These complications do not
arise in the minimal GM VFNS adopted by CTEQ, as already mentioned in the
previous subsection.)

For the neutral current DIS $F_{2}$ structure function, the above-mentioned
ambiguities can be see as follows:
\begin{equation}
\begin{array}{ccc}
& \mathrm{below} & \mathrm{above} \\
\rule{0 em}{3 ex} \mathrm{LO} & \frac{\alpha_{S}}{4\pi }C_{2,Hg}^{\mathrm{-}%
,1}\otimes g^{n_{f}} & C_{2,HH}^{+,0}\otimes (h+\bar{h}) \\
\mathrm{NLO} & \biggl(\frac{\alpha _{S}}{4\pi }\biggr)^{2}(C_{2,Hg}^{-,2}%
\otimes g^{n_{f}}+C_{2,Hq}^{-,2}\otimes \Sigma ^{n_{f}}) & \frac{\alpha _{S}%
}{4\pi }(C_{2,HH}^{+,1}\otimes (h+\bar{h})+C_{2,Hg}^{+,1}\otimes g^{n_{f+1}})
\\
\mathrm{NNLO} & \biggl(\frac{\alpha _{S}}{4\pi }\biggr)^{3} 
\sum_{i}C_{2,Hi}^{-,3}\otimes f_{i}^{n_{f}}&\biggl(\frac{\alpha _{S}}{4\pi }%
\biggr)^{2}\sum_{j}C_{2,Hj}^{+,2}\otimes f_{j}^{n_{f}+1},%
\end{array}
\label{mrsttab}
\end{equation}%
with obvious generalization to even higher orders. This means that switching
directly from a fixed order with $n_{f}$ active quarks to fixed order with $%
n_{f}+1$ active quarks leads to a discontinuity in $F_{2}^{H}(x,Q^{2})$. As
with the discontinuities in the ZM-VFNS already discussed this is not just a
problem in principle -- the discontinuity is comparable to the errors on
data, particularly at small $x$. The TR scheme, defined in \cite%
{trvfns1,trvfns2}, and all subsequent variations, try to maintain the
particular ordering in each region as closely as possible. For example at LO
the definition is
\begin{eqnarray}
F_{2}^{H}(x,Q^{2}) &=&\frac{\alpha _{S}(Q^{2})}{4\pi }%
C_{2,Hg}^{-,1}(Q^{2}/m_{H}^{2})\otimes g^{n_{f}}(Q^{2})  \nonumber \\
&\rightarrow &\frac{\alpha _{S}(m_{H}^{2})}{4\pi }C_{2,Hg}^{-,1}(1)\otimes
g^{n_{f}}(m_{H}^{2})+C_{2,HH}^{+,0}(Q^{2}/m_{H}^{2})\otimes (h+\bar{h}%
)(Q^{2}).  \label{trlo}
\end{eqnarray}%
The $\mathcal{O}(\alpha _{S})$ term is frozen when going upwards through $%
Q^{2}=m_{H}^{2}$. This generalizes to higher orders by freezing the term
with the highest power of $\alpha _{S}$ in the definition for $%
Q^{2}<m_{H}^{2}$ when moving upwards above $m_{H}^{2}$. Hence, the
definition of the ordering is consistent within each region, except for the
addition of a constant term (which does not affect evolution) above $%
Q^{2}=m_{H}^{2}$ which becomes progressively less important at higher $Q^{2}$%
, and whose power of $\alpha _{S}$ increases as the order of the
perturbative expansion increases.

This definition of the ordering means that in order to define a GM VFNS at
NNLO \cite{nnlovfns} one needs to use the $\mathcal{O}(\alpha _{S}^{3})$
heavy-flavour coefficient functions for $Q^{2}\leq m_{H}^{2}$ (and that the
contribution will be frozen for $Q^{2}>m_{H}^{2}$). This would not be needed
in a ACOT-type scheme. As mentioned above, these coefficient functions are
not yet calculated. However, as explained in \cite{nnlovfns}, one can model
this contribution using the known leading threshold logarithms \cite%
{hfthreshold} and leading $\ln (1/x)$ terms derived from the $k_{T}$%
-dependent impact factors \cite{hfsmallx}. This results in a significant
contribution at small $Q^{2}$ and $x$ with some model dependence. However,
variation in the free parameters does not lead to a large change.\footnote{%
It should be stressed that this model is only valid for the region $%
Q^{2}\leq m_{H}^{2}$, and would not be useful for a NNLO FFNS at all $Q^2$
since it contains no information on the large $Q^{2}/m_{H}^{2}$ limits of
the coefficient functions. A more general approximation to the $\mathcal{O}%
(\alpha _{S}^{3})$ coefficient functions could be attempted, but full
details would require first the calculation of the $\mathcal{O}(\alpha
_{S}^{3})$ matrix element $A_{Hg}$. This more tractable project is being
investigated at present \cite{NNNLOme}.}

The above discussions focused on $F^H_2$; but they mostly apply to $F_L$ as
well. We only need to mention that, with the adoption of the SACOT
prescription for heavy-quark initiated contributions (i.e.\ using the ZM
version of the Wilson coefficient), $F^H_L$ vanishes at order $\alpha_s^0$
as it does in the TR prescriptions.(This zeroth order coefficient function
does appear in some older GM VFNS definitions.) According to the MRST/MSTW
convention, the order $\alpha_s^1$ term of $F_L$ (both light and heavy
flavour) counts as LO, and so on, whereas In the CTEQ convention each
relative order is a power of $\alpha_S$ lower.

The general procedure for the GM VFNS for charged-current deep inelastic
scattering can work on the same principles as for neutral currents, but one
can produce a \textit{single} charm quark from a strange quark so $\chi
=x(1+m_{c}^{2}/Q^{2})$. However, there is a complication compared to the
neutral current case because the massive FFNS coefficient functions are not
known at $\mathcal{O}(\alpha _{S}^{2})$ (only asymptotic limits \cite{nlocc}
have been calculated). These coefficient functions are needed in a TR-type
scheme at low $Q^{2}$ at NLO, and for any GM VFNS at all $Q^{2}$ at NNLO.
This implies that we can only define the TR scheme to LO and the ACOT scheme
to NLO. However, known information can be used to model the higher order
coefficient functions similarly to the TR scheme definition to NNLO for
neutral currents. A full explanation of the subtleties can be found in \cite%
{MSTW08}.

\subsubsection{Scheme variations \label{sec:nnlo}}

The inclusion of the complete GM VFNS in a global fit at NNLO first appeared
in \cite{MSTW}, and led to some important changes compared a previous NNLO
analysis, which had a much more approximate inclusion of heavy flavours
(which was explained clearly in the Appendix of \cite{MRSTNNLO}). There is a
general result that $F_{2}^{c}(x,Q^{2})$ is flatter in $Q^{2}$ at NNLO than
at NLO, as shown in Fig.\thinspace 4 of \cite{MSTW}, and also flatter than
in earlier (approximate) NNLO analyses. This had an important effect on the
gluon distribution. As seen in Fig.\thinspace 5 of \cite{MSTW}, it led to a
larger gluon for $x\sim 0.0001-0.01$, as well as a larger value of $\alpha
_{S}(M_{Z}^{2})$, both compensating for the naturally flatter evolution, and
consequently leading to more evolution of the light quark sea. Both the
gluon and the light quark sea were $6-7\%$ greater than in the MRST2004 set
\cite{MRST04} for $Q^{2}=10,000\mathrm{GeV^{2}}$, the increase maximising at
$x=0.0001-0.001$. As a result there was a $6\%$ increase in the predictions
for $\sigma _{W}$ and $\sigma _{Z}$ at the LHC. This would hold for all LHC
processes sensitive to PDFs in this $x$ range, but would be rather less for
processes such as $t \bar t$ pair production sensitive to $x \geq 0.01$.
This surprisingly large change is a correction rather than a reflection of
the uncertainty due to the freedom in choosing heavy flavour schemes and
demonstrates that the MRST2004 NNLO distributions should now be considered
to be obsolete.

\begin{figure}[tbp]
\centering  \includegraphics[width=0.5\textwidth,clip]{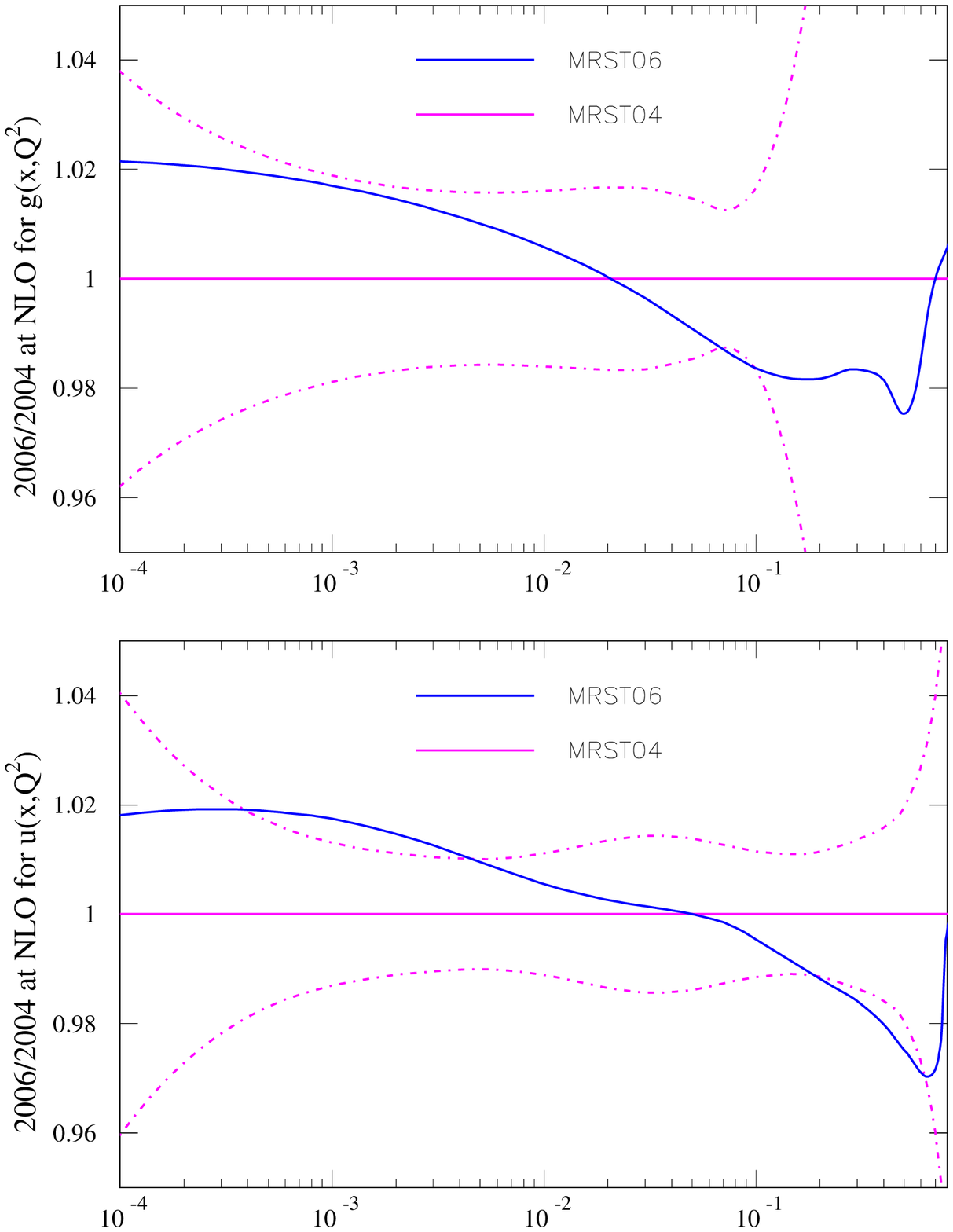}
\caption{\textsf{A comparison of the unpublished ``MRST2006 NLO'' parton
distributions to the MRST2004 NLO distributions. In order to illustrate the
significance of the size of the differences, the uncertainty on the MRST2001
distributions is used for the 2004 distributions.}}
\label{fig:06to04}
\end{figure}

To accompany the MRST 2006 NNLO parton update there is an unofficial
\textquotedblleft MRST2006 NLO\textquotedblright\ set, which is fit to
exactly the same data as the MRST2006 NNLO set. By comparing to the 2004
MRST set one can check the effect on the distributions due to the change in
the prescription for the GM VFNS at NLO without complicating the issue by
also changing many other things in the analysis. The comparison of the up
quark and gluon distributions for the ``MRST2006 NLO'' set and the MRST2004
NLO set, i.e. the comparable plot to Fig.\,5 of \cite{MSTW} for NNLO, is
shown in Fig.\,\ref{fig:06to04}. As can be seen it leads to the same trend
for the partons as at NNLO, i.e. an increase in the small-$x$ gluon and
light quarks, but the effect is much smaller -- a maximum of a $2\%$ change.
Also, the value of the coupling constant increases by $0.001$ from the 2004
value of $\alpha_S(M_Z^2)=0.120$. From momentum conservation there must be a
fixed point and this is at $x \sim 0.05$. Hence, $W, Z$ and lighter particle
production could be affected by up to $2-3\%$, and very high mass states by
a similar amount, but final states similar in invariant mass to $t \bar t$
will be largely unaffected. Hence, we can conclude that the change in our
choice of the heavy-flavour coefficient function alone leads to changes in
the distributions of up to $2\%$, and since the change is simply a freedom
we have in making a definition, this is a theoretical uncertainty on the
partons, much like the frequently invoked scale uncertainty. Like the
latter, it should decrease as we go to higher orders.

\subsection{Comparisons \label{sec:compare}}

We have tried to make clear that both the CTEQ and the MRST/MSTW approaches
are consistent with the PQCD formalism with non-zero heavy quark masses \{{$%
m_{H} $}${\}}$. In this sense, they are both \textquotedblleft
valid\textquotedblright . In addition, they both adopt certain sensible
practises, such as the numerically significant rescaling-variable approach
to correctly treat final-state kinematics (ACOT-$\chi $), and the
calculationally simplifying SACOT prescription for the quark-parton
initiated subprocesses. These common features ensure broad agreement in
their predictions. This is borne out by the fact that global QCD analyses
carried out by both groups show very good agreement with all available hard
scattering data, including the high-precision DIS total inclusive cross
sections and semi-inclusive heavy flavor production cross sections; and that
the predictions for higher energy cross sections at LHC for the important
W/Z production process agree rather well in the most recent versions of
these analyses \cite{cteq66,MSTW}.\footnote{%
Some apparent worrying discrepancies in the predictions for the W/Z
cross-sections at LHC between \cite{cteq65} and \cite{MRST04} have been
superseded by the recent analyses.} Comparisons of experiment for the
abundant data on total inclusive cross sections (and the associated
structure functions) with theory are well documented in the CTEQ and
MRST/MSTW papers. \ Here we only show the comparison of the recent H1 data
sets on cross sections for charm and bottom production \cite{H1botstruc} to
the latest CTEQ and MSTW calculations. This figure illustrates the general
close agreement between the two calculations. (Also, see below.)
\begin{figure}[h]
\includegraphics[width=1\textwidth,clip]{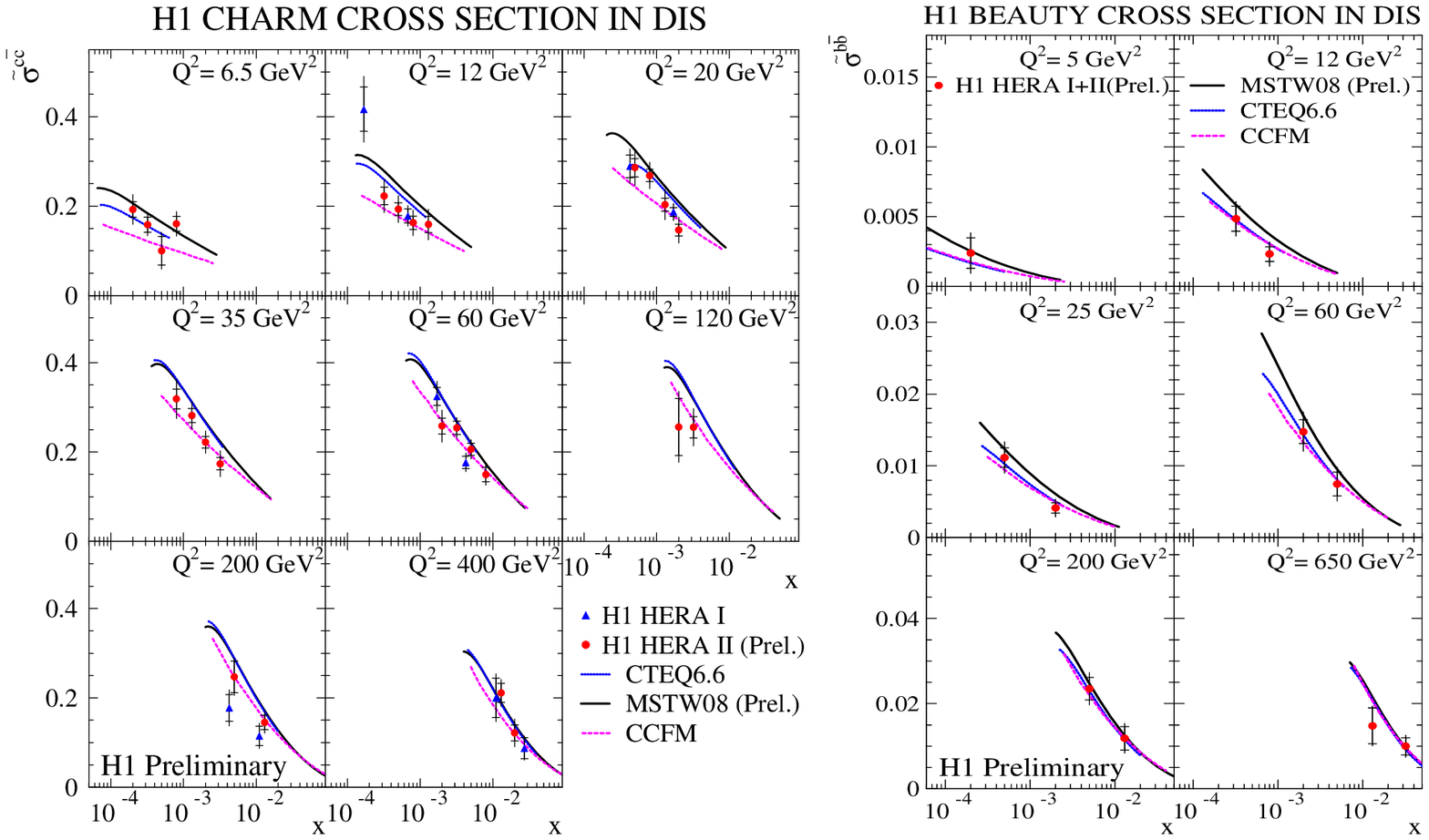}
\caption{\textsf{Comparison of the predictions for $\tilde{\protect\sigma}^{c%
\bar{c}}(x,Q^{2})$ and $\tilde{\protect\sigma}^{b\bar{b}}(x,Q^{2})$ compared
to preliminary data from H1.}}
\label{fig:h1bottom}
\end{figure}

Because the main source of the differences between the two implementations
arise from the different conventions adopted for organizing the perturbative
calculation, it is impossible to make a direct (or clear-cut) comparison
between the two calculations. By staying with the conventional
order-by-order formulation, the CTEQ approach has all the simplicities of
the minimal GM VFNS. With the alternative LO/NLO/NNLO organization, the
MRST/MSTW approach includes specifically chosen higher-order terms at each
stage of the calculation for different physical quantities (e.g.\thinspace $%
F_{2}^{\mathrm{tot}},F_{L}^{\mathrm{tot}},F_{2}^{\mathrm{H}}$, in Secs.\,\ref%
{sec:mrstF2}) with their associated Wilson coefficients (e.g.\thinspace
Eqs.\thinspace \ref{mrsttab},\ref{trlo}). The choices are a matter of taste
because, with the same Wilson coefficients (with heavy quark mass) available
in the literature (such as \cite{transitionme1, transitionme2}), both
analyses can be extended to the appropriate order, and they should contain
the same information. So far, MRST/MSTW has carried out their analyses to
one order higher than CTEQ. In practice, we have seen one comparison of the
\textquotedblleft NLO\textquotedblright\ predictions of the two approaches
in Fig.\,\ref{fig:h1bottom} that shows remarkable general agreement with
each other, and with experimental data. Some expected differences at small-$%
x $, due to the higher order term included in the MRST/MSTW calculation are
present. Compared to experimental data, the CTEQ curves seem to give a
slightly better description of data in this region of difference; but this
should not be taken seriously in view of the above discussions. We intend to
make a more quantitative study of the differences between the alternative
formulations of a GM VFNS and ZM VFNS in a future publication.

\section{Use of Parton Distribution Functions \label{sec:use}}

Some commonly asked questions in the user community for PDFs are along the
lines of: (i) Which available PDF set is most appropriate for my particular
calculation? and (ii) If PDF set A was obtained using scheme A (say,
ZMVFNS/GMVFNS-MSTW/GMVFNS-CTEQ) do I have to use the same scheme A for my
Wilson coefficients (otherwise my calculation would be inconsistent)?
Whereas it is impossible to answer all such questions at once, the following
observations should provide useful guidelines toward the appropriate
answers. Foremost, it is important to bear in mind that in the perturbative
approach, all calculations are approximate; hence the goodness of the
approximation is the most (or only) relevant consideration. Any fast, or
absolute, rules or prescriptions would be misguided.

\medskip \noindent \textbf{*} For applications at very high energy scales,
e.g.\thinspace most LHC processes, it is perfectly fine to use the ZM
formulae for the hard-scattering coefficient \emph{irrespective of the
choice of PDF sets }(see below), since the ZM Wilson coefficients are good
approximations to the GM ones (valid to $\mathcal{O}(M^{2}/Q^{2}$) where $M$
represents the typical mass in the relevant parton subprocess---heavy quarks
or other produced particles), and the ZM coefficients are much simpler and
\emph{much more} readily available.

On the other hand, for applications involving physical scales $Q\sim $ $%
\mathcal{O}(M)$, such as comparison to precision DIS data at HERA, it is
important both to use GM Wilson coefficients, and to ensure that these are
consistent with those adopted in generating the PDF set to be used in the
calculation.

\medskip \noindent \textbf{*} For the global analyses that yield the PDF
sets, it matters whether the ZM VFNS or GM VFNS scheme is used in the
calculation, since a substantial fraction of the input DIS data are in the
region where $Q$ is not very large compared to the heavy quark masses $%
m_{c,b}$ (the top quark does not play a significant role in these analysis).
Thus, the ZM-VFNS and GM VFNS PDFs can differ in some $x$-range, even if
they agree quite well in general (cf.\thinspace \cite{cteq65}). For example,
the widely used CTEQ6.1 (ZM-VFNS) and the most recent CTEQ6.5/CTEQ6.6 (GM
VFNS) PDF sets both give excellent fits to the available data, yet the
differences (mainly around $x\sim 10^{-3}$) are enough to lead to a $6\%$
shift in the predictions for cross sections for $W,Z$ and similar mass
states at the LHC. Higher mass final states are much less affected.

The above differences arise from two sources: (i) the treatment of
final-state counting (Sec.\thinspace \ref{sec:sumb}) and phase space
(Sec.\thinspace \ref{sec:rescaling}); and (ii) mass effects in the Wilson
coefficients. The first is numerically significant for reasons explained in
those sections, and it can potentially be removed to produce an improved ZM
VFNS (Sec.\thinspace \ref{sec:improveZM}).

\medskip \noindent \textbf{*} The differences between PDFs obtained using
different GM VFNS implementations, such as those by CTEQ and MSTW groups
discussed in the main part of this review, are much smaller than those
between the ZM and GM VFNS. This is because the treatments of final states
are similar, and the differences in the Wilson coefficients are much reduced
also. The current NLO predictions on $W/Z$ cross sections at LHC by the CTEQ
and MSTW groups, for instance, are within $2\%$ \cite{Watt08}.

\medskip \noindent \textbf{*} What about single-flavor (say, $n_{f}$) FFNS
PDFs that are commonly believed to be needed for FFNS calculations,
such as for heavy flavor production processes? We would like to point
out, perhaps surprisingly to many readers, that: (i) with the advent of
GM VFNS PDFs, \emph{the FFNS PDFs are not in principle needed} for
consistency; and (ii) the use of $n_{f}$-flavor FFNS PDFs in a $n_{f}$%
-flavor calculation is much \emph{less reliable} than using the GM VFNS (if
the latter is available). The reasons for these assertions are fairly easy
to see, as we now explain.

First of all, as we emphasized in Sec.\thinspace \ref{sec:suma}, the GM
VFNS is, by definition, a composite scheme that \emph{is} the
$n_{f}$-FFNS within the region of validity of the latter. In principle
one \emph{can} use the GM VFNS PDFs in the FFNS calculations within the
region where the FFNS is reliable. (In practice this range of validity
(in energy scale $\mu $) extends up to several times $m_{H}$, cf.\
second to last paragraph of Sec.\thinspace \ref{sec:suma}.) Secondly,
since any given $n_{f}$-FFNS has only a limited range of validity
(Sec.\thinspace \ref{sec:suma}), the global analysis used to determine
any $n_{f}$-FFNS PDF set is inherently a compromise. This compromise is
likely to be a fairly bad one for two reasons. Firstly, the limited
range of validity implies that only a fraction of the data used in the
global analysis can be legitimately applied. If one excludes all the
data outside of the region of validity of the theory (not an
easily-defined region), the constraining power of the analysis would
greatly suffer. If, instead, one includes all the points in the
analysis anyway, the PDFs will compensate, much like the case of the
fit using the basic ZM VFNS. This can result in a good comparison to
data (as in the ZM VFNS \cite{Alekhin}), but this is potentially
misleading since the compensation is caused by the wrong physics. In
either of the cases, the PDFs resulting from a fit using the FFNS will
be unreliable. Secondly, Wilson coefficients in the FFNS only exist for
the DIS process beyond LO, hence the ZM approximation to $n_{f}$-FFNS
must be used. We note, although this second point is shared by current
GM VFNS analyses, the ZM VFNS approximation to GM VFNS is a much better
approximation than that of ZM FFNS to $n_{f}$-FFNS. (For instance, for
collider jet data sets, the ZM 3- or 4-flavor calculation would be
way-off the correct one. This is not a problem for the GM VFNS case.)
These inherent problems motivated an alternative approach to FFNS PDFs
in \cite{MRSTFFNS}: rather than performing a (imperfect) FFNS global
fit, one simply generates them by fixed $n_{f}$-flavor QCD evolution
from a set of initial PDFs obtained in an existing (bona fide) GM VFNS
global analysis! Because of the different QCD evolution, however, the
PDFs will be different from the original GM VFNS ones crossing heavy
flavor thresholds; and the fits to the global data will correspondingly
deteriorate, particularly for the high precision HERA data sets at
higher $Q^{2}$. Thus, these PDFs deviate from truth in a different way.
The relative merit between this approach and the conventional FFNS
global fits is difficult to gauge because there are no objective
criteria for making the assessment.

Returning to the original question that started this bullet item, we
can summarize the options available to match PDFs with a FFNS
calculation such as HQVDIS  \cite{HQVDIS} for heavy quark production:
(i) conventional FFNS PDFs (CTEQ, GRV), suitably updated if necessary
\cite{GRV2}; (ii) PDFs generated by FFNS evolution from GM VFNS PDFs at
some initial scale $Q_{0}$ (MSTW \cite{MRSTFFNS}, but also can easily
be done with CTEQ); or, (iii) simply use the most up-to-date GM VFNS
PDFs (MSTW, CTEQ) for all $Q$. For reasons discussed in the previous
paragraphs, each option has its advantages and disadvantages. (i) and
even (ii) are theoretically self-consistent, while (iii) is not, e.g.
it opens up the akward question of how many flavours to use in the
definition of $\alpha_S$. However, the PDFs in (iii) are intrinsically
much more accurately and precisely determined. Hence, in practical
terms it is not obvious which would be most ``correct''.\footnote{Although it
is certainly better to use a current GM VFNS set of PDFs than an
out-of-date FFNS set.} The choice reduces to a matter of taste, and for
some, of conviction. The differences in results, obtained using these
options, should not be too large, since they are mostly of one order
higher in $\alpha_s$; and, in an approximate manner, they define the
existing theoretical uncertainty.  In principle, an approach that
combines the advantages of all three, hence could work the best, would
be to use PDFs obtained in the GM VFNS, but with the transition scale
$\mu_T$ (Sec.\,\ref{sec:sumb}) set at a much higher value than $m_H$
for each heavy flavor threshold.  But this option is rather cumbersome
to implement (as has been hinted in Sec.\,\ref{sec:sumb}), hence has
not been done.

\medskip \noindent \textbf{*} There exists another class of applications,
involving multiple-scale processes, such as heavy flavor production at
hadron colliders with finite transverse momentum $p_{T}$ or in
association with $W/Z$ or Higgs, for which PQCD calculations are more
complex than the familiar one-hard-scale case, as implicitly assumed
above. Since these processes can play an important role in LHC, there
has been much discussions, and controversies, in recent
literature about the various approaches that may be applied \cite%
{CampbellRev}. Both the GM VFNS \cite{HamburgGrp} and FFNS approaches have
been advocated \cite{Higgs}. The problem is complex, generally because more
than one kind of potentially large logarithms occur in these problems, and
they cannot be effectively controlled all at once with some suitable choice
of scheme. A detailed discussion is outside the scope of this paper,
although our remark about the FFNS PDFs above could be helpful (and relieve
some of the anxieties expressed in the literature).

All in all, for general applications, taking into account all the
considerations above, the modern GM VFNS PDF sets are clearly the PDFs
of choice.

\section{Intrinsic Heavy Flavour\label{sec:IC}}

Throughout the above discussions we have made the assumption that all heavy
quark flavour is generated from the gluon and lighter flavours through the
perturbative QCD evolution, starting from the respective scale $\mu =m_{H}$.
This is usually referred to as the \emph{radiatively generated heavy flavor}
scenario. From the theoretical point of view, this is reasonable for heavy
flavors with mass scale ($m_{H}$) very much higher than the on-set of the
perturbative regime, say $\sim 1$ GeV. Thus, while this assumption is
usually not questioned for bottom and top, the case for charm is less
obvious. In fact, the possibility for a non-negligible \emph{intrinsic charm}
(IC) component of the nucleon at $\mu =Q\sim m_{c}$ was raised a long time
ago \cite{Brodsky}; and interests in this possibility have persisted over
the years. Whereas the dynamical origin of such a component can be the
subject of much debate, the phenomenological question of its existence can
be answered by global QCD analysis: do current data support the IC idea, and
if so, what is its size and shape? This problem has been studied recently by
a CTEQ group \cite{CTEQintcharm}, under two possible scenarios: IC is
enhanced at high values of $x$ (suggested by dynamical models such as \cite%
{Brodsky}), or it is similar in shape to the light-flavor sea quarks
(similar to, say, strange). They found that current data do not tightly
constrain the charm distribution, but they \emph{can} place meaningful
bounds on its size. Thus, while the conventional radiatively generated charm
is consistent with data, IC is allowed in both scenarios. For the
model-inspired (large-$x$) case, the size of IC can be as large as $\sim 3$
times that of the crude model estimates, though comparison to the EMC $%
F_2^{c}$ data \cite{EMC} imply contributions somewhat smaller \cite{MRST98}.
\ If such an IC component does exist, it would have significant impact on
LHC phenomenology for certain beyond SM processes. \ For the sea-like IC
case, the bound on its size is looser (because it can be easily interchanged
with the other sea quarks in the global fits); its phenomenological
consequences are likewise harder to pin-point.

From a theoretical point of view, intrinsic heavy flavour and GM VFNS
definitions were discussed in \cite{thornedw}. Allowing an intrinsic
heavy quark distribution actually removes the redundancy in the
definition of the coefficient functions in the GM VFNS, and two
different definitions of a GM VFNS will no longer be identical if
formally summed to all orders, though they will only differ by
contributions depending on the intrinsic flavour. Consider using
identical parton distributions, including the intrinsic heavy quarks,
in two different flavour schemes. The heavy-quark coefficient functions
at each order are different by $\mathcal{O}(m_{H}^{2}/Q^{2})$. This
difference has been constructed to disappear at all orders when
combining the parton distributions other than the intrinsic heavy
quarks, but will persist for the intrinsic contribution. The intrinsic
heavy-flavour distributions are of $\mathcal{O}(\Lambda
_{QCD}^{2}/m_{H}^{2})$, and when combined with the difference in
coefficient functions the mass-dependence
cancels leading to a difference in structure functions of $\mathcal{O}%
(\Lambda _{QCD}^{2}/Q^{2})$. It has been shown \cite{collins} that for a
given GM VFNS the calculation of the structure functions is limited in
accuracy to $\mathcal{O}(\Lambda _{QCD}^{2}/Q^{2})$. Hence, when including
intrinsic charm, the scheme ambiguity is of the same order as the best
possible accuracy one can obtain in leading twist QCD, which is admittedly
better than that obtained from ignoring the intrinsic heavy flavour (if it
exists) as $Q^{2}$ increases above $m_{H}^{2}$. It is intuitively obvious
that best accuracy will be obtained from a definition of a GM VFNS where all
coefficient functions respect particle kinematics. In fact, the most recent
CTEQ and MSTW prescriptions would provide identical contributions to the
structure functions from the same intrinsic charm parton distribution.

\noindent

\paragraph{Acknowledgements}

We thank Matteo Cacciari for his unrelenting efforts to bring about
this review on behalf of the Heavy Flavor Workshop of the HERALHC
Workshop. We thank our collaborators in CTEQ and MRST/MSTW for
collaborations which underlies much of the content of this paper. WKT
would like especially to acknowledge the insight provided by John
Collins on the theoretical foundation of PQCD with heavy quarks. We
also thank Pavel Nadolsky, Fred Olness, Ingo Schienbein, Jack Smith and
Paul Thompson for reading the manuscript and making useful comments
that brought about improvements in the presentation of the paper.

The work of WKT is supported by the National Science Foundation (USA) under
the grant PHY-0354838.

{\raggedright
\bibliographystyle{heralhc}
\bibliography{heralhc}
}

\end{document}